\documentclass[preprints,article,accept,moreauthors,pdflatex,10pt,a4paper]{mdpi}
\newcommand{\cor}{_0} 	      
\newcommand{\crit}{_\text{c}} 

\newcommand{\EQ}{\begin{equation}}
\newcommand{\EN}{\end{equation}}
\newcommand{\EQA}{\begin{eqnarray}}
\newcommand{\ENA}{\end{eqnarray}}
\newcommand{\Sec}[1]{Section~\ref{#1}}

\newcommand{\Eq}[1]{Eq.~(\ref{#1})}

\newcommand{\bra}[1]{\langle #1\rangle}

\newcommand{\mean}[1]{\overline{#1}}

\newcommand{\emf}{\mathcal{E}}          

\newcommand\deriv[2]{\displaystyle\frac{\partial #1}{\partial #2}}

\newcommand\sfrac[2]{{\textstyle{\frac{#1}{#2}}}}
\def\vec#1{\ensuremath{\mathchoice{\mbox{\boldmath$\displaystyle#1$}}
		{\mbox{\boldmath$\textstyle#1$}}
		{\mbox{\boldmath$\scriptstyle#1$}}
		{\mbox{\boldmath$\scriptscriptstyle#1$}}}}
	
\newcommand{\AAA}{{\vec{A}}} 
\newcommand{\bb}{{\vec{b}}}
\newcommand{\BB}{{\vec{B}}}

\newcommand{\EE}{{\vec{E}}}

\newcommand{\FF}{{\vec{F}}}

\newcommand{\jj}{{\vec{j}}}
\newcommand{\JJ}{{\vec{J}}}
\newcommand{\kk}{{\vec{k}}}

\newcommand{\rr}{{\vec{r}}}

\newcommand{\vv}{{\vec{v}}}
\newcommand{\VV}{{\vec{V}}}

\newcommand{\xx}{{\vec{x}}}

\newcommand{\oo}{\vec{\omega}}

\newcommand{\nab}{{{\nabla}}}

\newcommand{\Prm}{\text{Pr}_\text{m}}
\newcommand{\Rey}{\text{Re}}          
\newcommand{\Rm}{R_\text{m}}        

\newcommand{\cm}{\,\text{cm}}    
\newcommand{\km}{\,\text{km}}    
\newcommand{\kpc}{\,\text{kpc}}  

\newcommand{\s}{\,\text{s}}      
\definecolor{webgreen}{rgb}{0,.5,0}
\definecolor{webbrown}{rgb}{.6,0,0}
\definecolor{purple}{rgb}{0.5,0,.5}
\definecolor{chocolate}{rgb}{0.82, 0.41, 0.12}
\newcommand{\ks}[1]{{#1}}

\index{gyration!frequency|see{Larmor frequency}}
\index{gyration!radius|see{Larmor radius}}
\index{material!derivative|see{Lagrangian derivative}}
\index{total!derivative|see{Lagrangian derivative}}

\usepackage[normalem]{ulem}
\firstpage{1} 
\pubvolume{xx}
\issuenum{1}
\articlenumber{5}
\pubyear{2018}
\copyrightyear{2018}
\history{Received: date; Accepted: date; Published: date}

\usepackage{float}
\usepackage{pstricks}

\Title{From primordial seed magnetic fields to the galactic dynamo}

\Author{Kandaswamy Subramanian $^{1,\dagger,\ddagger}$\orcidA{}*}

\AuthorNames{Kandaswamy Subramanian}

\address[1]{%
$^{1}$ \quad IUCAA, Post Bag 4, Ganeshkhind, Pune 411007, India; kandu@iucaa.in}

\corres{Correspondence: kandu@iucaa.in; Tel.: +91-20-25604101}

\abstract{The origin and maintenance of coherent magnetic fields in the Universe is reviewed
with an emphasis on the possible challenges that arise in their theoretical understanding.
We begin with the interesting 
possibility that magnetic fields originated at some level from the early universe.
This could be during inflation, the electroweak or the quark-hadron phase transitions. 
These mechanisms can give rise to fields which could
be strong, but often with much smaller coherence scales than galactic scales. Their
subsequent turbulent decay decreases their strength but increases their coherence.
We then turn to astrophysical \ks{batteries which can generate}
seed magnetic fields. Here the coherence scale can be large, but the field
strength 
\ks{is} 
generally very small. These seed fields need to be further amplified and maintained
by a dynamo to explain observed magnetic fields in galaxies. Basic ideas behind both small and large-scale turbulent dynamos are outlined.  The small-scale dynamo may help understand the first magnetization 
of young galaxies, while the large-scale dynamo is important for the generation of fields with scales larger than 
\ks{the} stirring \ks{scale}, 
as observed in nearby disk galaxies. The current theoretical 
challenges that turbulent dynamos encounter and their possible resolution are discussed. }


\begin{document}


\section{Introduction}
The universe is magnetized, right from the Earth, the Sun and other stars to disk galaxies, galaxy clusters and perhaps \ks{also} 
the intergalactic medium \ks{(IGM)} in voids.  In nearby disk galaxies, 
magnetic fields are observed to have both a coherent component 
of order a few micro Gauss, ordered on scales 
of a few to ten kilo parsecs (kpc) and 
a random component with scales of \ks{parsecs to} 
tens of parsecs \citep{F10,Beck16,HBGM08}.
In these galaxies, both stars and the gas in the 
interstellar medium (ISM), are in a thin disk supported against gravity by 
their rotation. It is not clear what is the strength and structure of magnetic 
fields in the other major type of galaxies, the ellipticals.
\ks{This is perhaps related to the fact that normal ellipticals have
much lower active star formation and lack the requisite
cosmic ray electrons for producing significant synchrotron emission.} 
There is tentative evidence that even young galaxies, which are several billion years younger than the Milky Way host
ordered micro Gauss strength magnetic fields \cite{Bernet+08,Farnes+14,Mao+17}.
Magnetic fields of similar
strengths and coherence are detected even in the hot plasma filling the most massive collapsed
objects in the universe, rich clusters of galaxies \cite{GF04}.
There is also indirect evidence for a lower limit of order $10^{-16}$G to
the magnetic field contained in the intergalactic medium of large scale void regions between galaxies \citep{NV10,TVN11} 
\ks{(see however \citep{BCP12}).
This strength refers to a coherence scale of a Mpc, and the field needs
to be stronger if the coherence scale is smaller.}  
The origin and maintenance of cosmic magnetism is an outstanding question of modern astrophysics.
We focus here on galactic magnetic fields and trace their origin and maintenance from the early to the present
day universe.

Magnetic fields of the observed strength need to be constantly maintained against turbulent decay, 
the turbulence either being self generated by the Lorentz force or driven by other forces. 
This is done by electromagnetic induction due to
motions in a preexisting magnetic field. Such motions can 
induce an electric field with a curl which by Faraday's law, can
maintain the magnetic field. 
The resulting evolution of the magnetic field B is governed by the induction equation,
\begin{equation}
\label{Induction1}
\deriv{\BB}{t}=
\nab\times\left(\VV\times\BB-\eta\nabla\times\BB\right),
\end{equation}
where $\VV$ is the fluid velocity and $\eta$ the resistivity of the plasma.
The first term in the induction equation describes the electromagnetic 
induction (the generation of electric field in a conductor moving across 
magnetic field), whereas the second term is responsible for its diffusion and 
resistive decay. If $\eta\to0$, the magnetic flux through any surface moving with 
the fluid remains constant. The relative importance of induction versus resistance is measured
by the dimensionless magnetic Reynolds number $\Rm= vl/\eta$, where $v$ and $l$ are typical values for the fluid
velocity and length scales respectively. For inter-stellar turbulence, 
\ks{adopting}
$v\sim 10 \km \s^{-1}$ \citep{DBB06}, $l \sim 100$ pc \ks{\citep{ACR81}} 
and 
\ks{Spitzer value for the resistivity} 
$\eta \sim 10^7 \cm^2 \s^{-1}$, 
as applicable to ionized plasma
at a temperature $T \sim 10^4$ K, we have $\Rm\sim 3\times 10^{19} \gg 1$.
From \Eq{Induction1} we see that one needs at least a seed magnetic field to be present
before induction can amplify it. It turns out that most ideas of seed field generation
lead to magnetic fields which are much smaller than observed. They need to be then amplified and
maintained, a process called the dynamo. We review ideas for both these
aspects. 

\section{Early Universe origin}

Seed magnetic fields could be a relic from the early Universe, arising 
during the inflationary epoch or in a later phase transition, when the electroweak symmetry is
broken or when quarks gather into hadrons (for reviews see \cite{DN13,S16}). 
Indeed, if the evidence for 
weak, 
femto Gauss magnetic fields in the void regions is firmed up, an early universe 
mechanism would provide a natural explanation. Such a possibility can
also help \ks{probing} 
the physics of the early universe.  In this section, we
use the natural system of units in which the Planck constant, speed
of light and the Boltzmann constant are equal to 1.

In the expanding universe all length scales increase proportional
to the expansion factor $a(t)$. 
\ks{Thus if the magnetic flux is frozen, the field strength at a time $t$ 
decreases or redshifts as $B(t) \propto 1/a^2(t)$.}
\ks{Neglecting the effects of dissipation, its}
energy density \ks{then decreases as}
$\rho_B(t)=B^2(t)/(8\pi) \propto 1/a^4(t)$. 
The energy density of 
cosmic microwave background radiation (CMB),  $\rho_\gamma(t)$, 
a relic of the hot 'big bang' \ks{beginning} of the universe,
\ks{also decreases with expansion in the same manner. This implies the
approximate constancy of}
the ratio $r_B=\rho_B(t)/\rho_\gamma(t)$ 
(
Approximate \ks{as particle annihilation at} 
certain epochs, 
can increase \ks{$\rho_\gamma(t)$}). 
\ks{This motivates characterizing the strength of }
the primordial field 
with either \ks{$r_B$} 
or  
$B_0$ \ks{the field strength at the present epoch,}
as a function of 
the scale $L$ over which the field is averaged.
 A 
\ks{value of} $B_0 \sim 3.2 \mu $G 
\ks{corresponds to the field having the same energy density
as the CMB today,}
or $r_B=1$. Observations of CMB anisotropy or structure formation
lead to upper limits of $B_0$ at the nano Gauss levels assuming
nearly scale invariant magnetic spectrum 
\cite{DN13,TSS14,PCSF15,PlanckB15,S16}.

\subsection{Generation during Inflation}
\label{geninf}

The seeds for structures we see in the universe are thought to have originated during the inflationary epoch, 
from amplification of quantum vacuum fluctuations in the scalar field driving the rapid accelerating expansion of
the universe. 
Inflation does have several useful features to generate coherent seed magnetic fields as well \cite{TW88}. 
First, the rapid expansion during inflation stretches small scale 
wave modes to very large correlation scales corresponding to galaxies and larger. Second, such expansion
dilutes any pre-existing charge densities to be negligible.
Then 
\ks{the conductivity of the universe is negligible,
there is no constraint from conservation of}
magnetic flux \ks{and one can generate magnetic fields} 
from a zero field. The idea is then to excite quantum 
fluctuations of the \ks{vacuum state} of \ks{the}
electromagnetic (\ks{perhaps} more correctly hypermagnetic)  
field, when a given mode is
within what is known as the Hubble radius, which then 
transits to random classical fluctuations as the mode is stretched well
beyond the Hubble scale. Subsequently, when the universe reheats generating 
charge particles, the electric field is shorted and damped to zero, while the
magnetic field part of what once was an electromagnetic wave is frozen into the
resulting plasma.

\ks{This idea however faces} one major difficulty. 
The \ks{conventional} 
electromagnetic (EM) 
action $S_{EM}$ is left invariant under a conformal 
transformation of the metric ($g_{\mu \nu}$) 
given by $g^*_{\mu \nu} = \Omega^2 g_{\mu \nu}$.
\ks{Moreover,} 
the geometry of the 
Friedmann-Robertson-Walker (FRW) 
expanding universe itself transforms to its flat space version under a suitable
conformal transformation. Then 
Maxwell equations \ks{and consequently} 
the electromagnetic wave equation 
transform to \ks{what obtains in flat space-time.}
\ks{In such a case,}
\ks{EM} wave fluctuations 
\ks{cannot be amplified}
in a FRW universe. 
The \ks{electromagnetic} field \ks{still decays}
with expansion as $1/a^2(t)$, which is very drastic during inflation.
Therefore significant inflationary magnetogenesis requires a 
mechanism for breaking conformal invariance of the electromagnetic action,
\ks{so that the decrease of the field becomes milder, to say}
$B \sim 1/a^{\epsilon}$ with 
$\epsilon \ll 1$.
A \ks{variety of models where such a behaviour can obtain}
have been \ks{suggested},
one of them being to
couple a scalar field $\phi$ (perhaps the inflaton responsible for driving inflation) 
to the EM action as $S= f^2(\phi) S_{EM}$ during inflation \citep{TW88,Ratra92}. 
It turns out that in this model, one gets a scale
invariant spectrum of magnetic fields for $f\propto a^2$ or $f\propto a^{-3}$, 
with a present day amplitude \cite{S16}
\begin{equation}
B_0 \sim 0.6 \times 10^{-10} {\rm G} 
\left(\frac{H}{10^{-5} M_{pl}}\right)\; .
\label{Bstrength}
\end{equation}
Here $H$ is the Hubble expansion rate in energy units during inflation and 
$M_{pl}$ the Planck energy. 
Thus, \ks{for specific evolutionary behaviour} 
of the coupling function $f$, \ks{strong enough} 
fields 
can 
be \ks{generated}.

\subsubsection{Constraints and Caveats}

A number of constraints and caveats arise in models of 
inflationary magnetogenesis. First, a time dependent
coupling $f$ in front of the EM action implies that electric
fields and magnetic fields evolve differently. For example in the 
model with $f \propto a^{-3}$ electric fields increase rapidly with time
even though the magnetic field remains almost constant. Then
its energy density can begin to exceed the inflaton energy density
causing a back reaction problem \cite{MY08,FM12}. This does not happen in the model
with $f\propto a^2$. However in the latter model
the function $f =f_i (a/a_i)^2$ increases
greatly during inflation, from its initial value of $f_i$ at $a=a_i$.
When the interaction
of the EM field with charged particles is taken \ks{into} account, 
the value of $f$ at the end of reheating, $f_0$, 
\ks{will} renormalize the 
electric charge \ks{from} 
$e$ to 
$e_N = e/f_0^2$.
\ks{Suppose we require the charge to take the present day value
at the end of inflation, i.e $f_0 =1$. Then}
$f_i \ll 1$ \ks{initially and thus}
$e_N  = e/f_i^2 \gg e$ \ks{at early times}. \citet{DMR09} argued 
that the theory is not trustable 
\ks{in this case as the EM field}  
is in a strongly coupled regime.
Alternatively, suppose one started with a weakly coupled theory where
$f_i \sim 1$. Then $f_0\gg f_i$ \ks{by} the end of inflation 
and so
the renormalized charge $e_N \ll e$. 
\ks{When inflation ends, the interaction of the electromagnetic} field 
\ks{with the charges will then be}
 extremely weak.
A third potential problem raised by \cite{KA14} is that the
\ks{creation of charged particles by the}
generated electric fields, 
due to the Schwinger effect, 
\ks{can increase the} conductivity \ks{so much that} 
magnetic field generation
\ks{freezes}.

We have built models which 
attempt to address these issues by having a rising $f$ during inflation followed
by a decreasing $f$ until reheating, but now predict a blue magnetic field spectrum 
$d\rho_B/d\ln k \propto k^4$ \ks{($k$ is the comoving wavenumber)} 
and require a low energy scale of inflation and reheating 
\cite{Ram17,Ram18}. 
The spectrum is cut-off at the Hubble wavenumber 
of reheating. 
The field is also helical when one adds a parity breaking piece to the EM action \cite{Ram18}.
In this case the field orders itself considerably as it decays (see below). 
We find that a scenario with reheating at
a temperature of 100 GeV leads to present day 
field strengths of order $B_0 = 4 \times 10^{-11}$ G with a coherence
scale of $70$ kpc. 

\subsection{Generation during phase transitions}
\label{PhaseB}

As the universe expands and cools from very high temperatures, it goes through
the \ks{electroweak} (EW) phase transition (at $T=T_c\sim 100$ GeV) 
and the quark-hadron (QCD) phase transition (at $T_c\sim 150$ MeV). 
Significant magnetic field generation can take
place in these phase transitions, especially if they are of first order.
\ks{In this case, the transition to} 
the new phase 
\ks{occurs in bubbles nucleating in}
the old phase. \ks{These bubbles} 
expand and collide \ks{with each other} until
\ks{the universe transits completely to the new phase.}
In these bubble collisions battery effects can
operate to generate a seed magnetic field which is further amplified by a dynamo due to the
turbulence generated during bubble collisions \cite{Hogan83}.
\ks{The consequences of such a picture has been
studied for both the}
EW phase transition \citep{Baym_etal96}
and the QCD phase transition \citep{QLS89,Sigl_etal97}.
More subtle effects have also been considered invoking gradients in the Higgs field during
EW phase transition \cite{vachaspati91}, linking baryogenesis with magnetogenesis \cite{Cornwall97,V01},
or using the chiral anomaly of weak interactions \cite{BRS12,B2017}. A brief review of some of these effects is
given in \cite{S16}.
The properties the magnetic field generated in all these
models is uncertain but $\rho_B$ 
can be
a few percent of $\rho_\gamma$. 
\ks{The} coherence scale \ks{of the field} 
can \ks{be as small as a}
few tens of the thermal de-Broglie
wavelength $1/T$ \ks{upto} a \ks{significant} fraction 
$f_c$ of the Hubble scale.
For the EW phase transition, which occurs at a
temperature of about $100$ GeV, the proper Hubble scale is
of order a cm, and thus the comoving coherence scales
will then be of order $10^{15} f_c$ cm \cite{S16}.
For the QCD phase transition, which would occur at 
a temperature of $T\sim 150$ MeV, the Hubble radius is
$\sim 6.4 \times 10^5$ cm, and the comoving coherence scale
is of order $(f_c/3)$  pc.
Moreover, the present-day strength for a magnetic field which has
say a fraction $r_B=0.01$ 
is $B_0 \sim 0.3 \mu$G.

\subsubsection{Magnetic field evolution in the early universe}

The small-scale magnetic fields generated in these phase transitions
or in the inflationary models with blue spectrum \cite{Ram17,Ram18}, 
are strong enough to
drive decaying magnetohydrodynamic (MHD) turbulence \cite{BEO96,BJ04}. 
The magnetic field energy density then decreases faster than the $(1/a^2)$ 
dilution due to expansion. However, the field 
coherence scale simultaneously also increases with the decay. 
Note that in the radiation dominated universe, MHD equations
reduce to their flat space-time version provided one uses a
conformally transformed fields, for example $B_* = B a^2$, 
conformal time $\tau =\int dt/a(t)$ and comoving 
spatial coordinate $\xx = (\rr/a(t))$ ($\rr$ is the proper spatial coordinate). 
Moreover, the plasma in the early universe is an excellent electrical conductor, 
but its viscosity increases whenever the mean free path of a particle species
(like the neutrino or photon) grows to be comparable with the coherence
scale of motions. In epochs when viscosity dominates, the peculiar 
velocity induced by the Lorentz force becomes damped
and hence does not in turn distort the field, freezing its evolution. 
In all other epochs, the Lorentz force induced velocity leads to
decaying MHD turbulence.

In the case of decay of fluid turbulence in flat space-time, a general feature is 
that of preservation of large scales (larger than the coherence scale) 
during the decay, and then the evolution of energy and coherence scale
depends on the energy spectrum on such large scales \cite{Davidson04}.
The case of nonhelical magnetic field
decay appears to be more complicated. 
Numerical simulations find that
the comoving magnetic energy density, $E_M \propto (B_*^2/8\pi)$ decays slower 
than for pure hydro turbulence, as $E_M \propto \tau^{-1}$ and undergoes an
inverse transfer of energy with the coherence
scale $L_c(\tau)$ increasing as $\tau^{1/2}$ \cite{BKT15,RB17}.
If the field is fully helical, magnetic helicity conservation
constrains the decay and further slows it down to $E_M \propto \tau^{-2/3}$
while $L_c$ increases faster as $\tau^{2/3}$ \citep{CHB01,BJ04}. 
\ks{If the field is} partially helical 
\ks{its} decay \ks{is to begin with} as \ks{in the nonhelical case but} 
conserving helicity. \ks{This makes}
the field 
\ks{eventually} fully helical \ks{after which}
they decay \ks{more slowly like in} 
the fully helical case. 
For the radiation
dominated universe with $a(t) \propto t^{1/2}$, we also have $\tau \propto t^{1/2}\propto a(t)$.
Thus a power law decay in conformal time is still a power law decay in physical time (though slower).
When matter starts dominating, the transformation law to the flat-space time 
MHD equations are different \cite{BJ04} and the relevant time co-ordinate becomes $\tilde{\tau}=\int dt /a^{3/2}$.
Since the expansion $a(t)\propto t^{2/3}$ in the matter dominated era, $\tilde{\tau}\propto \ln(t)$ and any power
law decay for the comoving magnetic field in $\tilde{\tau}$ becomes only a logarithmic decay in real time. 
Therefore turbulent decay of the field almost freezes after matter domination. 

\subsubsection{Predicted field strengths and coherence scales}

These ideas have been put together by several authors
\cite{BJ04,DN13,BKT15,KBT16,S16,Ram17,Ram18} to estimate \ks{$B_0$ and $L\cor$},the 
\ks{field strength} 
and coherence scale \ks{respectively}
\ks{at the present epoch,}
of magnetic fields which can undergo nonlinear evolution.
First, a general constraint relating 
$B_0$ and  
 $L\cor$ can be found from the following criterion, that 
the field decays to a strength where the 
Alfv\'en crossing time across $L\cor$ 
\ks{equals} the \ks{present} 
age of the universe \cite{BJ04}. This gives
\begin{equation}
B_0 \approx 5 \times 10^{-12} {\rm G} \left(\frac{L\cor}{{\rm kpc}}\right).
\label{BJlimit}
\end{equation}
\ks{The field strength $B_0$ itself can be estimated}
assuming the \ks{scaling laws for} turbulent decay 
\ks{starting} from generation 
(\ks{when} $a=a_g$, $T=T_g$) 
to end of \ks{matter} radiation \ks{equality} 
($a=a_{eq}$, $T \sim 1$ eV). \ks{As discussed above, we assume}
the comoving field strength \ks{changes negligibly thereafter}.
For nonhelical fields assuming the possibility of inverse transfer 
\cite{BKT15}, gives
$B_0 = (a_{eq}/a_g)^{-1/2}B_g$, where $B_g$ is the
comoving magnetic field $B_*$ at 
generation. 
The scale factor ratio can be related to the temperature ratio
using entropy conservation 
during the radiation era. This gives 
$aTg^{1/3}$ being constant with expansion, where
\ks{we denote by}
$g$ 
the 
degrees of freedom \ks{of relativistic particles}.
\ks{We assume} 
$g \sim 100$ at the \ks{epoch of} generation \ks{and}  
$g\sim 4$ at the \ks{matter-radiation} equality. \ks{This} 
gives 
\EQ
B_0
\sim 6 \times 10^{-13} \left(\frac{r_B}{0.01}\right)^{1/2} T_{100}^{-1/2} \ {\rm G},
\quad L\cor \sim 0.1 \kpc \left(\frac{r_B}{0.01}\right)^{1/2} T_{100}^{-1/2} 
\label{nonhelets}
\EN
where $T_{100} = T_g/(100 \ {\rm GeV})$ and the coherence scales is obtained 
from using $B_0$ in \Eq{BJlimit}.
For the
\ks{case where the field is only} partial helical,
with \ks{$h_g$ the} initial helical fraction, 
we have $B_0= B_g (a_{eq}/a_h)^{-1/3} (a_h/a_g)^{-1/2}
= B_g (a_{eq}/a_g)^{-1/3} (a_g/a_h)^{1/6}$. \ks{Here} $a_h$ is expansion
factor and 
$\tau_h$ \ks{the corresponding epoch,} 
when the \ks{decay of the} field 
\ks{makes it} 
fully helical. 
\ks{The initial helical fraction is defined as the ratio of 
initial helicity $H_g$ by the maximal helicity $H_{max}$ 
for a given energy, which for a peaked magnetic spectrum is 
$H_{max} \simeq B_g^2 L_c(\tau_g)$ \citep{BJ04}.}
\ks{We note} 
that the initial helicity $H_g$ 
is \ks{nearly} 
conserved while energy decays.
\ks{Therefore the} fractional
helicity subsequently scales as
$h \simeq H_g/(E_M(\tau) L_c(\tau)) = 
h_g (\tau/\tau_g)^{1/2}$ and becomes unity when 
$(\tau_g/\tau_h) = (a_g/a_h) = h_g^2$ or when $(a_g/a_h)^{1/6} =h_g^{1/3}$.
Hence $B_0= B_g (a_{eq}/a_g)^{-1/3} h_g^{1/3}$ and
putting in numbers, 
\EQ
B_0 
\sim \ks{10^{-10}} 
\left(\frac{r_B}{0.01}\right)^{1/2} T_{100}^{-1/3} h_g^{1/3} \ \ks{ {\rm G}},
\quad L\cor \sim 20 \kpc \left(\frac{r_B}{0.01}\right)^{1/2} T_{100}^{-1/3} h_g^{1/3}.
\label{helest}
\EN
\ks{The above}
estimates agree reasonably with
\ks{that of \citet{BJ04} from their detailed
simulations of the magnetic field decay.}

Thus primordial magnetic fields surviving from
the early universe could account for the 
\ks{lower limits to}
magnetic field in voids 
\ks{coming from the $\gamma$-ray
observations of blazars emitting in the TeV energies. Their} 
\ks{field} strengths and coherence scales 
\ks{could even be such as} 
to influence other physical processes
in the universe.

\ks{It is important to mention the following caveat with
the $\gamma$-ray constraints. For this we recall
how the constraint is obtained. Firstly it is argued that
high energy TeV photons from 
a blazar interact with eV photons in the intergalactic space
to produce a beam of relativistic electron-positron ($e_{\pm}$) pairs, after
travelling a distances of order tens of Mpc, typically into
the void regions. This $e_{\pm}$ beam inverse 
Compton scatters the ambient CMB photons to GeV energies such that
one should see a Gev $\gamma$-ray halo around every TeV blazar, which
is not detected. This null result can be explained if the beam gets
sufficiently spread out due to deflection of electrons and
positrons in opposite directions by an intergalactic
magnetic field, which leads to the lower limit on such fields. 
However, there is ongoing debate as to whether the
$e_{\pm}$ beam traversing through the intergalactic medium, 
loses its energy due to plasma
instabilities at a rate faster than the inverse Compton rate 
\citep{BCP12,SIS12,ME13,DN13,Chang14,KKS15,Brod+18}. 
In such a case, one would not see a GeV halo around the
blazar even if there were no magnetic fields in the voids
and consequently no lower limit on the intergalactic field 
would be obtained.
Irrespective of the final outcome of this debate,
the fact that one can potentially probe such a weak intergalactic magnetic 
field from $\gamma$-ray astronomy is very exciting. Of course, 
such a field need not be primordial, but could 
arise from the pollution of magnetic fields
from galactic outflows \citep{BVE06,SSS18}, but the volume filling
factor of such outflows is uncertain. 
}

\ks{An} important challenge for 
\ks{magnetogenesis} scenarios \ks{involving phase transitions, 
is the requirement that they ideally be of first order.}
The EW or QCD phase transition are first order 
only in extensions to the standard
model of particle physics \citep{SchStu09,GSW05,HKPS07}. In
the standard model, 
\ks{the EW and QCD}
phase transitions are 
'crossover' transitions,
\ks{with}
thermodynamic variables 
\ks{changing} 
continuously \ks{but significantly in} 
a narrow range of temperature \ks{around the critical temperature 
$T_c$}
\citep{CFH98,Aoki06}. 
\ks{Magnetogenesis} models \ks{which involve}
phase transitions \ks{of first order} in the early universe and/or
which generate strong magnetic fields with a blue power spectrum, like
in the inflationary magnetogenesis 
models of \citep{Ram17,Ram18}, can
lead to a significant stochastic gravitational wave background.
This can be probed by space gravitational
wave detectors like LISA in the future \cite{CDS09,CF18,Roper+19,Ram19}.

\section{Astrophysical batteries and seed magnetic fields}

The Universe is charge neutral but positive and negative charged particles have different 
masses \ks{-} a 
feature which is at the root of many astrophysical battery mechanisms.

\subsection{Biermann batteries}

For example suppose a pressure gradient is applied to a fully ionized hydrogen plasma.
Pressure depends on number density and temperature and if these are the same for electrons and protons,
the force on these fluid components will also be identical. 
However the electrons, being much lighter than protons, will be accelerated 
much more than the protons. This relative acceleration leads to an 
electric field, $\EE = -\nab p_e/e n_e$, which couples back positive and negative charges so that 
they move together, \ks{obtained by equating} 
the \ks{electron} pressure gradient 
$-\nab p_e$ 
with the electric force $-e n_e \EE$. Here $n_e$, $p_e = n_e k T$ 
and $T$ are respectively 
the number density, pressure and temperature of the electron fluid, and we have assumed that 
protons are much more massive and so do not move. 
If this thermally generated electric field has a curl, from Faraday's law, magnetic fields can
grow from zero. Adding this electric field in Ohm's law and taking the curl gives a modified
induction equation,
\begin{equation}
\frac{\partial \BB}{\partial t} =
\nab\times \left(\VV \times\BB -\eta\nab \times \BB\right)
-\frac{c k_{\rm B}}{e} \frac{\nab n_e}{n_e}\times\nab T.
\label{modb}
\end{equation}
We see that \Eq{modb} now contains a {\it source} term such that 
magnetic fields can be generated from initially zero fields.
This source is nonzero if the density and temperature gradients, 
$\nab n_e$ and $\nab T$, are not parallel to each other, and
the resulting battery effect is known as the Biermann battery.
It was first proposed as a mechanism for the
generation of stellar magnetic fields \citep{Bier50,mestelrox},
but has subsequently found wide applications to the cosmological
context as well \citep{SNC94,KCOR97}.

For example during reionization of the universe by star bursting galaxies and quasars, 
the temperature gradient is normal to the ionization front.
However density gradients are determined by arbitrarily laid down density fluctuations, 
which will later collapse to form galaxies and clusters, and which need not be correlated 
to the source of the ionizing photons. The source term in \Eq{modb} is then nonzero
and magnetic fields coherent on the scale of the density fluctuations, or 
galactic and larger scales can grow. This will be amplified further 
during the collapse to form galaxies and one expects a seed magnetic field in galaxies
$B \approx 10^{-21}$ G \citep{SNC94}. Direct numerical simulations of cosmic 
reionzation \cite{GFZ00} have confirmed such
a scenario, and find a magnetic field ordered on Mpc scales, with a mass weighted
average $B \sim 10^{-19}$ G at a redshift of about 5.

The Biermann battery can also operate in oblique 
cosmological shocks which arise during the formation of galaxies 
and large scale structures to generate magnetic fields \cite{KCOR97}.
For partially ionized hydrogen,
with uniform ionization fraction $\chi$ and all species having the
same temperature, $p_e = \chi p/(1+\chi)$ and $n_e = \chi \rho/m_p$.
Here $p$ is the total fluid pressure.
Defining $\vec{\Omega}_B = e\BB/m_pc$, \Eq{modb} reduces to
the same form as the induction equation but now for $\vec{\Omega}_B$
with a source term $(\nab p \times\nab \rho)/(\rho^2(1 + \chi))$.
This source term, without the extra factor $-(1 + \chi)^{-1}$,
corresponds to the baroclinic term in the vorticity
equation for $\vec{\Omega} = \nab \times \VV$, where \ks{the} 
Lorentz force is neglected.
Thus provided viscosity and resistivity are
neglected, $\vec{\Omega}_B (1 +\chi)$ and $-\vec{\Omega}$
satisfy the same equation,  and if they were both zero initially, they will always be
equal later, i.e $e\BB/m_p = - \vec{\Omega}/(1 +\chi)$.
Taking the vorticity associated with spiral galaxies,
\EQ
\vert\BB\vert \approx 10^{-19} {\rm G} \left(\frac{\Omega}{10^{-15} \s^{-1}}\right).
\label{Bvort}
\EN
Direct numerical simulations were used by \citet{KCOR97} to calculate
the vorticity build up in structure formation shocks, which using \Eq{Bvort} 
then translates into a seed magnetic field of $B\sim 10^{-21}$ G in regions about to collapse into 
galaxies at redshift $z\sim 3$.
\subsection{Battery due to interaction with radiation}

\ks{The difference between the masses}
of positive and negative charges 
can also lead to battery effects 
\ks{when an ionized plasma interacts with radiation.}
\ks{Indeed} 
electrons are more strongly coupled with 
radiation than the protons, 
\ks{because}
the Thomson cross section for \ks{its} 
scattering off photons \ks{is larger, being} 
inversely \ks{proportional to} 
the mass of the \ks{charged} particle. 
Due to this photon-electron/proton scattering asymmetry,
during recombination, both vorticity and magnetic fields are generated in the
second order of perturbations. The strength of these seed fields are 
again of tiny $B \sim 10^{-30}$G on Mpc scales
up to $B \sim 10^{-21}$G at parsec scales \citep{GS05,MMNR05,TIOH05,KMST07}.
Moreover, during reionization of the universe, the radiative force from a
source is larger on electrons than protons, accelerating the electrons
more, again generating an electric field which couples them back together.
Due to the inhomogeneity of the intergalactic medium this electric field will have a curl
leading to magnetic field generation. This field is estimated to be 
between $10^{-23}-10^{-19}$ G on coherence scales between hundreds of
kpc to pc respectively \citep{DL15,D+17}.

\subsection{Plasma effects}
\label{plasma}

During cosmological structure formation, 
\ks{the infall kinetic energy 
of the intergalactic medium (IGM) is expected 
to be converted into thermal energy through many shocks.}
The densities in the IGM are small with 
$n \sim 2 \times 10^{-7} (1+z)^3 \cm^{-3}$,
and therefore \ks{Coulomb} collisions may not be strong enough 
to form these shocks,
\ks{and one may need other means for particle collisions.}
Plasma instabilities like the Weibel instability \citep{W59,F59}, 
which occur when there are
counter streaming plasma motions, generate small scale magnetic fields, which then
effectively scatter particles. The idea that these fields 
provide seed magnetic fields has been explored by several authors \citep{MSK06,LSWP09}.
Such a plasma instability has typical growth times
$\tau_p \sim (v/c)^{-1} (1/\ks{\omega_i})$, where $v$ is the upstream velocity, 
$\omega_i= (4\pi n_i e^2/m_i)^{1/2}$ the plasma frequency, and 'i' can 
represent electrons or ions, 
with coherence scales corresponding to the species skin depth $c/\omega_i$. These timescales
are so small even for ions, $\tau_p \approx 6 \times 10^2   v_2^{-1} n_{-5}^{-1/2}$ s,
compared to astrophysical timescales
that the instability would rapidly saturate. 
Here $v_{2} = v/(10^2 \km \s^{-1})$ \ks{of order $1-3$} is 
a typical inflow \ks{velocity} for galaxies \ks{which will have
velocity dispersions of the same order} and
$n_{-5} = n/(10^{-5} \cm^{-3})$ the IGM density 
at redshifts of $z\sim 4-5$, \ks{given its $(1+z)^3$ scaling}.

Particle in cell simulations show that saturation occurs when the field grows
to  a small fraction $\epsilon_B$ of the kinetic energy density of the 
inflowing plasma. Then the gyro radius of ions becomes 
smaller than the skin depth, whereby
particles will get strongly deflected and so not counter stream.
The resulting magnetic fields at saturation can be strong
with $B \sim  3 \times 10^{-9} G (\epsilon_B/10^{-3})^{1/2} v_2 n_{-5}^{1/2}$, but correlated on 
the very small ion-skin depth 
$10^{-8} n_{-5}^{-1/2}$ pc
\citep{KT08,CSA08}. The long time survival of this shock generated field 
is unclear. Moreover, averaged over galactic scales they can only provide a tiny seed
field for the dynamo (see \Sec{largeseed}).

\subsection{Seed fields from stars and \ks{active galactic nuclei (AGN)}}

A seed magnetic field for the galaxy can also be provided by 
ejection of stronger magnetic fields from stars and active galaxies
which have a much shorter dynamical time scale and form before
the bulk of the galactic interstellar medium gets magnetized
\citep{RSS86,Rees87,Rees05,Rees06}. These processes can give
fairly large seed magnetic fields of order a nano Gauss or larger.
Of course in this case the dynamo has to operate efficiently in
stars and AGN, and faces the challenges that we describe
later in this review. There is also the issue of how magnetized
plasma ejected from these objects is mixed with the originally 
unmagnetized interstellar medium in a protogalaxy, and how
this affects its strength and coherence scales.

\subsection{Large-scale seed magnetic field from small scale fields}
\label{largeseed}

In several contexts that we have discussed, the generated seed magnetic field 
even if strong, has a much smaller coherence scale than that of galaxies.
In order to estimate the seed this provides for the galactic dynamo one has
to determine the long wavelength (small wavenumber $k$) tail of the
corresponding 1-dimensional magnetic power spectrum $M(k)$. 
For hydrodynamic turbulence, both a 1 dimensional velocity
power spectrum $E(k) \propto k^2$ (called the Saffman spectrum) 
and $E(k)\propto k^4$ are possible \cite{Davidson04}.
It has been argued that $M(k) \propto k^4$ for the magnetic case
using $\nab\cdot\BB=0$ and the analyticity of the power spectrum \cite{DC03}. 
To elucidate the 
conditions required for this, we proceed as follows:

The magnetic correlation function in Fourier space $\hat{M}_{ij}(\kk)$
is the Fourier transform of the real space correlation function. 
Contracting the indices, assuming statistical isotropy and homogeneity, 
the 3-D \ks{magnetic spectrum} $M_{3d}(k)$ is given by
\begin{equation}
M_{3d}(k) = \frac12\int w(r) e^{i\kk.\rr} d^3r = 2\pi \int  \frac{d}{dr}\left[r^3M_L \right] 
\frac{\sin(kr)}{kr} dr =  2\pi \int  \frac{d}{dr}\left[r^3M_L \right] [ 1 - k^2r^2/6 + ...].
\label{m3dk}  
\end{equation}
Here we have used the fact that for \ks{a} 
statistically isotropic and homogeneous
magnetic field $w(r) = \mean{\bb(\xx)\cdot\bb(\xx+\rr)} = 1/r^2 d(r^3M_L)/dr$
and $M_L(r)$ is the longitudinal correlation function \cite{S99,BS05}.
The last step in \Eq{m3dk} has made a small $kr$ expansion of $\sin(kr)$.
The first term in the expansion in \Eq{m3dk} is $r^3M_L$ evaluated at infinity, and goes to zero
if $M_L$ falls off faster than $1/r^3$.
Then the next term dominates at small $k$, provided the resulting integral
is non zero, which it would be in general for $M_L(r)$ falling of
sufficiently rapidly. Then $M_{3d}(k) \propto k^2$ and so
the 1-d spectrum $M(k) \propto k^2M_{3d}(k) \propto k^4$.
On the other hand, if the magnetic field correlator $M_L(r)$ falls 
\ks{off as} $1/r^3$ due to
the persistence of long range correlations, then the first term
in the integral does not vanish and instead, goes to a constant.
Then $M_{3d}(k) \to {\rm constant}$ 
as $kr\to 0$, hence $M(k) \to k^2$. 
The first case would hold for example when the field is in 
randomly oriented magnetic field rings (or flux tubes),
while the latter case will \ks{be obtained} if one generates instead
randomly oriented current rings.
So both cases of $M(k) \propto k^2$ (random current rings) and $M(k) \propto k^4$
(random B flux rings) would seem possible depending on the origin of the field.
For a spectrum $M(k)\propto k^n$, the power per logarithmic interval in $k$ space
scales as $kM(k)\propto k^{n+1}$, and hence the magnetic field smoothed
over a volume of size $l=1/k$ scales as $B_l \propto l^{-(n+1)/2}$.

Suppose the field is coherent on a small scale $l$, has strength on this scale $B_l$, 
and the spectrum goes as $M(k) \propto k^n$ for $kl \ll1$, an estimate of the power on a 
large scale $L\gg l$ is given by $B_L \sim B_l (l/L)^{(n+1)/2}$. For example, in case of
the Weibel instability generated field of \Sec{plasma}, with $B_l \sim 3 \times 10^{-9}$ G
at $l \sim 3 \times 10^{10} \cm$, taking $L = 1 \kpc$, we get
$B_L \sim 10^{-25}$ G even for the $n=2$ case. On the other hand if
supernovae seed fields of $B_l \sim 10^{-6}$ G on scales of $100$ pc, on a larger galactic 
scale of say $L=3 \kpc$ the seed field would be $B_L \sim 6 \times 10^{-9}$ G for $n=2$ case
and $B_L \sim 2 \times 10^{-10}$ G, which are fairly strong
seed magnetic fields for a dynamo to act on. 


\section{Turbulent dynamos and their challenges}

Turbulence or random motions, which is prevalent in all systems from stars 
to galaxy clusters 
is thought to be crucial for amplification of seed magnetic fields to the
observed levels, a process called "turbulent dynamo". 
Turbulence is driven mostly by supernovae in the galactic 
interstellar medium \citep{ES04}, 
although during the formation of a galaxy by collapse from
the IGM, accretion shocks and flows along cold streams could also be important 
\citep{KH10,EB10,CDB10,LSSN13,MDST18}. 
In disk galaxies shear due to the differential rotation also plays
an important role in the dynamo amplification process.
Turbulent dynamos are conveniently divided into two classes, the fluctuation or small-scale
and mean-field or large-scale dynamos. This split depends respectively on whether the
generated field is ordered on scales smaller or larger than the scale of the turbulent motions. Here we briefly outline their
role in galactic magnetism focusing on the challenges that they present. Much of
our current understanding of these dynamos come from their
analysis using statistical methods or direct numerical simulations (DNS).
We shall focus more on some conceptual issues here.

\subsection{Fluctuation or small-scale dynamos}

The fluctuation dynamo is generic to sufficiently 
highly conducting plasma which hosts random motions, perhaps due to turbulence. 
First, in such plasma, magnetic flux through any area
moving with the fluid is conserved. Moreover, in any turbulent flow, fluid parcels random 
walk away from each other and so magnetic field lines get extended. 
Consider a flux tube with plasma of density $\rho$,  
magnetic field $B$, area of cross section $A$ and linking
fluid elements separated by length $l$. Flux conservation implies 
$BA = {\rm constant}$. Mass conservation in the flux tube gives
$\rho A l = {\rm constant}$, which implies $B/\rho \propto l$. 
Thus if $l$ increases
due to random stretching and $\rho$ is roughly constant,
then $B$ increases. This of course comes at the cost of 
$A \propto 1/\rho l\propto 1/B$
decreasing, the field being concentrated on smaller and smaller scales 
till resistivity becomes important 
\ks{at a scale $l_B$.} 
An estimate for this resistive scale 
gives $l_B \sim l\Rm^{1/2}$,
got by balancing the decay rate due to resistive diffusion, $\eta/l_B^2$, with 
growth rate due to random stretching $v/l$. 
Here $v$ and $l$ are the velocity and 
its coherence scale respectively of turbulent eddies.
As $\Rm$ is typically very large 
in astrophysical systems, the resistive scale $l_B\ll l$. 

What happens when resistive dissipation balances random stretching  
can only be addressed by a quantitative calculation.
The first such calculation was due to \citet{Kaz68}, who considered an
idealized random flow which is $\delta$-function correlated in time. For such a
flow one can write an exact evolution equation for the two-point magnetic correlator, which
has exponentially growing solutions, or is a dynamo, when $\Rm$ exceeds
a modest critical value $R\crit \sim 100$. 
The growth rate is a
fraction of the eddy turn over rate $v/l$, and at this kinematic stage, 
the field is shown to be concentrated on the scale $l_B$.
\ks{From the idealized Kazantsev model, it also turns out that 
$R\crit$ is larger and the growth slower 
for compressible flows compared to the incompressible case 
\citep{RK97,SChob+15,MMV19}.} 
\ks{Moreover, for Kolmogorov turbulence where the flow is multi-scale,
ranging from the outer scale to the small scales where
viscosity dominates, the fastest amplification is by the 
smallest supercritical eddy motions.
In the galactic ISM, the kinematic 
viscosity $\nu$ is typically much larger than the resistivity $\eta$,
and then growth would be expected to occur first due to dynamo action 
by the smaller viscous scale eddies \citep{KA92,S98}.} 
For the interstellar turbulence with \ks{an outer scale of 
turbulence}
$l\sim 100$ pc and velocities $v\sim 10 \km \s^{-1}$, we expect
$\Rm \gg R\crit$ and a growth time scale $l/v \sim 10^7$ yr 
\ks{even by the largest eddies}.
This \ks{time scale} is so much smaller than ages of even young high redshift 
galaxies, say a few times $10^9$ yr old, that the fluctuation
dynamo is expected to rapidly grow even weak seed magnetic fields to 
micro Gauss levels. 
\ks{Moreover, as smaller eddies can grow the field faster, 
significant amplification occurs even earlier.}
However as $l_B \ll l$, the field in the
growing phase is extremely intermittent and concentrated \ks{on} the
small resistive scales.  The big challenge is then whether
these fields can become coherent enough to explain for example
observations of the Faraday rotation inferred in young galaxies.

This growth of random magnetic fields due to the fluctuation dynamo 
has been verified by direct numerical simulations of driven turbulence, 
albeit in the idealized setting of isothermal plasma, for both subsonic
and supersonic flows \cite{HBD04,Schek04,CR09,BS13,PJR15,TCB12,F16,SBS18}).
Such simulations however have a modest 
values of $\Rm/R\crit \sim 10-20$. 
The basic expectations of the idealized Kazantsev 
model during the kinematic phase are qualitatively verified. 
The field grows exponentially and is concentrated initially
on the resistive scales. 
\ks{It is also found that the small-scale dynamo is less efficient 
for compressible compared to solenoidal forcing, 
as it generates less vorticity \citep{HBM04,Fed+11,F16}.
Importantly,} the DNS can now also follow the field 
evolution in to the nonlinear regime when Lorentz forces act to saturate the dynamo. 
By the time the dynamo saturates, 
the coherence length of the field increases to be a fraction
of order $1/3-1/4$ the scale of the driving, at least when the magnetic
Prandtl number $\Prm=\nu/\eta$ is of order unity 
\citep{HBD04,Eyink_etal13,CR09,BS13,PJR15}. 
\ks{These DNS have resolutions from $512^3$ upto $2048^3$.}
\ks{More modest resolution ($256^3$) DNS with  
large $\Prm$ but small fluid Reynolds number $\Rey$
found the magnetic energy spectrum to be still peaked
at the resistive scale $l_B$ even at saturation \citep{Schek04}.}
It is difficult to directly simulate the
case expected in the interstellar medium, of both a large $\Rm/R\crit$ and 
large \ks{$\Rey$}, as one then has to resolve both the widely
separated resistive and viscous dissipation scales.
\ks{Clearly the saturated state of the fluctuation dynamo
deserves further study, especially in this highly 
turbulent and $\Prm \gg1$ regime.}  

We have also directly determined Faraday rotation measures (RMs), in simulations
of the fluctuation dynamo with various values of $\Rm$, fluid Reynolds number $\Rey$ 
and up to rms Mach number of ${\cal M} = 2.4$ \cite{SSH06,BS13,SBS18}.
At dynamo saturation, for a range of parameters, we find an rms RM contribution 
which is about half the value expected if the field is coherent on the turbulent
forcing scale.  
\ks{Interestingly,}  
in subsonic and transonic cases,
the general sea of volume filling fields, 
\ks{dominates in determining the strength of RM}.
The rarer, strong field  structures, \ks{contribute only
about $10-20\%$ to the RM signal, indicating that
perhaps the coherence of the generated fields is associated
with more typical volume filling magnetic field regions.} 
However, \ks{when the turbulence is supersonic significant
contributions to the RM also comes from}
strong field regions as 
well as moderately over dense regions.
How exactly the field orders itself during saturation is at present 
an open problem.

\ks{One may wonder if magnetic reconnection is important for dynamo action. 
We note that the reconnection speed, depends inversely on the magnetic field 
strength even when it is efficient. Thus it would be too long compared
to the dynamo growth rate to be relevant during the kinematic stage
of the dynamo. It could however play a role once the field becomes
dynamically important. Some interesting aspects of a reconnecting
flux rope dynamo have been explored in \citep{BBSS09}. 
In nearly collisionless plasmas like galaxy clusters, plasma
effects could set transport properties even for weak fields 
and small scale dynamo action
in such a context is just beginning to be explored \citep{RCSV16,SK18}.}

Simulations of galaxy formation from cosmological initial conditions have also showed
evidence for amplification by the fluctuation dynamo, over and above 
the result of amplification by 
flux freezing during the compressive collapse to form the galaxy 
\cite{RT16,RT17,Pakmor17,Martin+18,Marinacci18}. 
\ks{One of the main limitations of such 
cosmological simulations is the resolution; that it will be very
difficult to capture both the galactic scale and the dissipative
scales, to predict correctly the rate of growth of magnetic energy
and the coherence scale of the saturated field.} 
Intriguingly, some of the direct simulations of SNe driven
turbulence, which have possibility of a multiphase medium, 
do not yet show a strong fluctuation dynamo \cite{Gressel08,Gent13,Bendre15},
although they do show large-scale dynamo action (except for \cite{Balsara04}).

All in all, one expects energy of random, intermittent magnetic fields 
to generically grow rapidly in the turbulent ISM of galaxies. This turbulence
could be driven by supernovae in disk galaxies. Galactic disks would then
host significant  fields, and a line of sight going through the
disk could have a significant RM \cite{BS13,SBS18}. This can partly
explain the statistical detection of excess RM in MgII absorption systems 
\cite{Bernet+08,Farnes+14}, which are thought to be associated with young galaxy disks at 
redshifts $z\sim1$. However, the abundance of these systems gives evidence
that the MgII absorption arises 
not only in line of sights through the disk, but also in extended gaseous halos 
\cite{Church05}. Thus one would need the halo
to be also magnetized and produce a significant RM.  This could occur
through outflows from the disk which also carry cold magnetized "clouds".
More work is required to firm up such a speculation.

\subsection{Mean-field or large-scale dynamos and galactic magnetism}

Remarkably, when turbulence is helical, magnetic fields on scales
larger than the coherence scale of the turbulence can be amplified. 
In any rotating, stratified system like the ISM of a disk galaxy
random motions driven by supernovae do become helical due to the Coriolis force, 
with one sign of helicity in the
northern hemisphere and the opposite sign in the southern hemisphere.
Such helical turbulent motions of the plasma draw out toroidal fields 
in the galaxy into a twisted 
loop generating poloidal components
(called the $\alpha$-effect). Differential rotation of the disk
shears \ks{the} radial component of the poloidal \ks{field} 
to generate back 
a toroidal component (the $\omega$-effect). These two can combine
to exponentially amplify the large-scale field provided \ks{that} 
the generation terms can overcome an extra resistivity
due to the turbulence. This is quantified  
by a dimensionless dynamo number being supercritical.  
Turbulent resistivity 
also allows the mean-field flux to be changed.

Quantitatively, in mean-field dynamo theory, the 
total magnetic field is split as $\BB = \mean\BB +\bb$, 
the sum of a mean (or the large-scale) field $\mean\BB$ 
and fluctuating (or the small-scale) field $\bb$.
A similar split of the velocity field gives $\VV =\mean\VV +\vv$.
The mean is defined by some form of averaging on scales larger
than the turbulence coherence scale, ideally but not necessarily
satisfying Reynolds rules for such averaging.
\ks{These rules are \citep{MY07}:
$\overline{\BB_1+\BB_2}=\mean\BB_1+\mean\BB_2$, 
$\overline{\mean\BB}=\mean\BB$, so $\overline{\bb}=0$, 
$\overline{\mean\BB\bb}=0$, 
$\overline{\mean\BB_1\;\mean\BB_2}=\mean\BB_1\;\mean\BB_2$ 
and averaging commutes
with both time and space derivatives.}
The induction equation \Eq{Induction1} then averages to give
\begin{equation}
\label{Induction_meanfield}
\deriv{\mean\BB}{t}
=\nab\times\left(\mean\VV\times\mean\BB+\vec{\emf}
-\eta\nab\times\mean\BB\right).
\end{equation}
Here a new term quadratic in the fluctuating fields arises,
the mean electromotive force (EMF),  $\vec{\emf}=\overline{\vv\times\bb}$.
To express this in terms of the mean fields themselves
presents a closure problem, even when Lorentz forces
are not yet important. The simplest such closure,
which is valid when the correlation time $\tau$ is
small compared to $l/v$  gives $\vec{\emf} = \alpha_K \mean\BB-\eta_t \nab\times \mean\BB$,
where the turbulent motions are also assumed to be isotropic.
Here $\alpha_K = -\sfrac13\tau\bra{\vv\cdot\oo}$ with $\oo=\nab\times\vv$, 
depending on the kinetic helicity of the turbulence 
and is the $\alpha$-effect mentioned above while 
$\eta_t=\sfrac13\tau\bra{\vv^2}$ is a turbulent diffusivity \ks{and} 
depends on
the kinetic energy of the turbulence. In disk galaxies 
we also have a $\mean\VV= r\Omega(r)\vec{\phi} $ corresponding 
to its differential rotation with frequency $\Omega$ along the toroidal 
direction $\vec{\phi}$. The mean-field dynamo equation~(\ref{Induction_meanfield}) 
with this form for $\vec{\emf}$ and $\mean\VV$, 
has exponentially growing solutions provided
a dimensionless dynamo number has magnitude
$D = \vert \alpha_0 S h^3/\eta_t^{2} \vert > D_{crit} \sim 6$ 
\cite{RSS86,Shukurov05,SS19}. 
Here $h$ is the disk scale height and $S=r d\Omega/dr$ the galactic shear,
$\alpha_0$ \ks{a} typical value of $\alpha$, and we have defined $D$ to be positive.
This condition can be satisfied in disk galaxies and the mean field 
typically grows on the rotation time scale, $\sim 10^8-10^9$ yr.
A detailed account of mean-field theory predictions for galactic dynamo theory
and its comparisons to observations is done by other authors in this volume.
We focus on the challenges for this general paradigm, in our view.

\subsubsection{Magnetic helicity conservation}

The first potential difficulty, which has already received
considerable attention, arises due to the conservation of
magnetic helicity in the highly conducting galactic plasma.
Magnetic helicity 
is usually defined as $H=\int_V \AAA\cdot\BB \ dV$ over a 'closed' volume $V$,
with $\AAA$ the vector potential satisfying $\nab\times\AAA =\BB$.
It is invariant under a gauge transformation $\AAA'=\AAA+\nab\Lambda$ 
only if the normal component of the field on the boundary to volume $V$ 
goes to zero. Magnetic helicity
measures the linkages between field lines \citep{berger_field84,Blackman15},
is an ideal invariant and is better conserved than total energy  in
many contexts, even when resistivity is included.
The mean-field dynamo works by generating poloidal from toroidal field
and vice-versa and thus automatically generates links between
these components, and thus a large-scale magnetic helicity.
To conserve the total magnetic helicity, corresponding oppositely signed
helicity must then be transferred to the small-scale field, which as we shall see
is done by the turbulent emf $\mean\emf$.

In fact, when helical motions writhe the toroidal field to generate a poloidal field, 
an oppositely signed twist must develop on smaller scales, 
to conserve magnetic helicity. For the same magnitude of
magnetic helicity on small and large scales, the Lorentz
force $(\JJ\times\BB)/c$ 
is generally stronger on small-scales (since $\JJ$ the current 
density has two more derivatives compared to the vector potential
which determines magnetic helicity). 
Thus Lorentz forces 
associated with this twist helicity can unwind the field 
while turbulent motions writhe it.  
According to closure models like 
\ks{the} Eddy damped quasi linear Markovian (EDQNM) approximation \cite{PFL76} 
or the $\tau$ approximation \cite{BF02,RKR03,BS05}, Lorentz forces then 
lead to an additional effective magnetic $\alpha$-effect, $\alpha_M = 
\frac13 \tau \overline{\jj\cdot\bb}/4\pi\rho$, with 
the total $\alpha=\alpha_K+\alpha_M$. The generated magnetic $\alpha_M$
opposes the kinetic $\alpha_K$ produced by the
helical turbulence and quenches 
the $\alpha$-effect and the dynamo, making it subcritical, 
much before the large-scale field grows strong enough
to itself affect the turbulence. For avoiding such quenching,
small-scale helicity must be shed from the galactic interstellar
medium. In principle resistivity can dissipate small-scale  magnetic helicity
but this takes a time longer than the age of the universe! 
For large-scale dynamos to work small-scale helicity must be lost more
rapidly, through magnetic helicity fluxes \cite{BF00,BS05,Blackman15}.

Magnetic helicity being a topological quantity, one may wonder how 
to define its density and its flux!  
A Gauge invariant definition of helicity density was given by \citet{SB06}
using the Gauss linking formula for the magnetic field 
\cite{moffatt69,berger_field84}. They proposed that
the magnetic helicity density $h$ of a random magnetic 
field $\bb$ is the density of correlated links 
of the magnetic field \cite{SB06}. 
This definition by construction involves only the random field $\bb$,
works if this field has a small correlation scale compared to the
system scale, and is closest to the helicity density defined
using the vector potential in Coulomb gauge.
An evolution equation can then
be derived for this density of helicity which now also
involves a helicity flux density $\mean\FF$ \citep{SB06},
\EQ\label{finhel}
\deriv{h}{t} + \nab\cdot\mean{\FF}
= -2\vec{\emf}\cdot\mean{\BB}-2\eta\overline{\nab\times\bb\cdot\bb}\,.
\EN
This equation involves transfer of magnetic helicity from
large to small scales by the turbulent emf along the
mean field ($ -2\vec{\emf}\cdot\mean{\BB}$ term), 
the  dissipation by resistivity ($-2\eta\overline{\nab\times\bb\cdot\bb}$) 
and the spatial transport by the helicity flux ($\nab\cdot\mean{\FF}$).
In the absence of such a flux,  and in the steady state we
see that $\vec{\emf}\cdot\mean{\BB}= -2\eta\overline{\nab\times\bb\cdot\bb}$
and so the emf along the field, which is important for the dynamo,
is resistively suppressed for $\Rm\gg1$. Even in the time dependent case, 
as the $\mean\BB$ builds up, $h$ also grows
and produces an $\alpha_M$ which cancels $\alpha_K$
to suppress the net $\alpha$ effect. 
In the presence of helicity fluxes however, $h$ can be transported out 
of the system allowing mean-field
dynamos to work efficiently \citep{BF00,SSSB06,SSS07}.

One such flux is simply advection of the gas and
its magnetic field out of the disk, i.e. $\mean\FF = h\mean\VV$ \cite{SB06,SSSB06}. 
 Several other types of helicity 
fluxes have been calculated like the Vishniac-Cho flux \ks{depending} 
on shear and the mean field \cite{VC01,BS05} and a
flux involving inhomogeneous $\alpha$ \cite{KMRS00}.
A diffusive flux $\mean\FF= - \kappa \nab h$ was postulated by 
\cite{KMRS02} and subsequently measured in DNS \cite{MCCTB10}. 
A new type of helicity 
flux which depends on purely an inhomogeneous random magnetic field and
rotation or shear has been worked out by \citet{Vishniac}, and could be potentially
important to drive a large-scale dynamo purely from random fields
in the galaxy, but has not yet
been studied in detail. 
Both the diffusive flux and the later Vishniac flux have been derived
from the irreducible triple correlator contribution to $\mean\FF$ by \cite{GS19} 
using a simple $\tau$-closure theory, but they also find several
other terms which cannot be reduced to either of these forms.
A detailed study of magnetic helicity
fluxes still remains one of the important challenges of the future. 

As an interesting application of these ideas,
\citet{CSS15} solved the mean-field dynamo equation incorporating
both an advective flux and a diffusive flux in \Eq{finhel}. 
Advection can be larger from the optical spiral, where star formation and galactic 
outflows are expected to be enhanced.
The helicity fluxes allow the mean-field dynamo
to survive, but stronger outflow along spiral arms led to
a relative suppression of mean field generation there and an
interlaced pattern of magnetic and gaseous arms as seen in the galaxy
NGC6946 \citep{BeHo96}. Interestingly a wide spread magnetic spiral
only results if the optical spiral is allowed to wind up and thus
here we are constraining spiral structure theory 
using magnetic field observations \cite{CSS13,CSS15}!
In another direction, the cosmic evolution of large-scale magnetic fields
during hierarchical clustering in the universe to form galaxies,
has also been extensively explored \cite{Luiz15,Luiz19}. 
 
\subsubsection{Mean-field dynamo in presence of the fluctuation dynamo}

We have discussed possibilities of both fluctuation and mean-field
dynamos in the turbulent interstellar medium. However random magnetic fields
due to the fluctuation dynamo grow much faster on time scale \ks{of} $10^7$ yr,
at least a factor 10 faster than the mean-field. 
Lorentz forces can then become important 
to saturate the field growth much before the mean field has
grown significantly. Will then these strong fluctuations make mean field
theory invalid? And can the large-scale field then grow at all? 
Earlier work \cite{S98} suggested that perhaps the intermittency
of the small-scale dynamo generated field on saturation still allows the
Lorentz force to be sub dominant in the bulk, and thus allow large-scale field growth. 
\citet{BSB16} examined this issue using
direct simulations of magnetic field amplification
due to fully helical turbulence in a periodic box, following
up earlier work on the kinematic stage by \cite{SB14}. Turbulence was
forced at about $1/4$ th the scale of the box, so that in principle
both scales smaller and larger than forcing can grow.
Initially all scales grow together as a shape invariant eigen function 
dominated by power on small-scales. This behaviour is akin to what
happens in fluctuation dynamos. But crucially on saturation of small
scales due to the Lorentz force, larger and larger scales continue to grow, and 
come to dominate due to the mean-field dynamo action. Finally  
system scale fields (here the scale of the box) develop
provided small-scale magnetic helicity can be efficiently removed,
which in this simulation is due to resistive dissipation.
\ks{Recent work by \citet{BSB19} 
in fact now finds evidence for two stages of
exponential growth, the sequential opertaion of both the 
small-scale dynamo, and
as it saturates, a quasi-kinematic large-scale dynamo, 
which is indeed exciting!}
This issue of how the small- and large- scale dynamos come to terms with each other 
deserves much more attention including a better analytic understanding.  

\section{Final thoughts}

We have traced briefly the generation of magnetic fields right from the early 
universe to their subsequent amplification by turbulent dynamos in the
\ks{later} universe. Several challenges remain to be addressed
in each of the processes that \ks{were} discussed. Apart from the issues already raised, early universe mechanisms 
need to be put in
the context of particular particle physics models. As far as the dynamo, their saturation
behaviour and how coherent the resulting fields become still raises intriguing questions.
The observational future appears bright.
\ks{A key objective of the} 
Square Kilometre Array (SKA)
\ks{is to elucidate}
the origin of cosmic magnetism.
The determination of a large number
of RMs and their modelling will likely yield rich dividend 
\cite{GBF04,Roy+16}. Of particular
interest will be to probe magnetic fields in the high redshift universe and 
the field in intergalactic filaments which could reflect more pristine
conditions. Surprisingly, Gamma-ray observations 
of TeV blazars have suggested lower limits at femto Gauss levels 
to the magnetic field in the IGM associated with large scale voids. Such
weak magnetic fields are difficult to detect by other techniques and
so it would be worthwhile to continue such studies.
Gravitational wave astronomy, especially the detection of a stochastic background, 
could also help to probe phase transitions and associated
magnetogenesis in the early universe.
Clearly study of cosmic magnetism will continue to be fascinating.

\acknowledgments{I thank Pallavi Bhat, Eric Blackman, 
Axel Brandenburg, Luke Chamandy, T. R. Seshadri, 
Ramkishor Sharma, Anvar Shukurov and Sharanya Sur for
enjoyable discussions on issues raised here.
I also thank Axel Brandenburg and 
Peter Davidson for very interesting correspondence 
which led to the discussion in
Section 3.5.}
\



\externalbibliography{yes}
\bibliography{seedgalr}



\end{document}